\documentclass[sigconf]{acmart}

\AtBeginDocument{%
  \providecommand\BibTeX{{%
    \normalfont B\kern-0.5em{\scshape i\kern-0.25em b}\kern-0.8em\TeX}}}

\copyrightyear{2021} 
\acmYear{2021} 
\setcopyright{acmcopyright}\acmConference[SIGIR '21]{Proceedings of the 44th International ACM SIGIR Conference on Research and Development in Information Retrieval}{July 11--15, 2021}{Virtual Event, Canada}
\acmBooktitle{Proceedings of the 44th International ACM SIGIR Conference on Research and Development in Information Retrieval (SIGIR '21), July 11--15, 2021, Virtual Event, Canada}
\acmPrice{15.00}
\acmDOI{10.1145/3404835.3463260}
\acmISBN{978-1-4503-8037-9/21/07}

\begin{document}
\fancyhead{}
\title{WTR: A Test Collection for Web Table Retrieval}

\author{Zhiyu Chen}
\email{zhc415@lehigh.edu}
\affiliation{%
  \institution{Lehigh University}
  \streetaddress{113 Research Drive (Building C)}
  \city{Bethlehem}
  \state{PA}
  \country{USA}
  \postcode{18015}
}

\author{Shuo Zhang}
\email{szhang611@bloomberg.net}
\affiliation{%
  \institution{Bloomberg}
  \city{London}
  \country{United Kingdom}
}

\author{Brian D.\ Davison}
\email{davison@cse.lehigh.edu}
\affiliation{%
  \institution{Lehigh University}
  \streetaddress{113 Research Drive (Building C)}
  \city{Bethlehem}
  \state{PA}
  \country{USA}
  \postcode{18015}
}

\renewcommand{\shortauthors}{Chen et al.}

\begin{abstract}
We describe the development, characteristics and availability of a test collection for the task of Web table retrieval, which uses a large-scale Web Table Corpora extracted from the Common Crawl. 
Since a Web table usually has rich context information such as the page title and surrounding paragraphs, we not only provide relevance judgments of query-table pairs, but also the relevance judgments of query-table context pairs with respect to a query, which are ignored by previous test collections. 
To facilitate future research with this benchmark, we provide details about how the dataset is pre-processed and also baseline results from both traditional and recently proposed table retrieval methods. Our experimental results show that proper usage of context labels can benefit previous table retrieval methods.

\end{abstract}




\maketitle

\section{Introduction}
Nowadays, tables have been used in various research tasks such as table-based question answering~\cite{sun2016table,chen2020hybridqa}, table-to-text generation~\cite{liu2018table,Chen2020FewshotNW}, column type annotation~\cite{chen2018generating,yi2018recognizing,hulsebos2019sherlock}, table retrieval and so on.
Among those tasks, table retrieval has obtained much attention recently in the IR community, witnessed by unsupervised methods~\cite{zhang2018ad,trabelsi2019improved,chen2020leveraging}, semantic feature-based methods~\cite{zhang2018ad,zhang2019table2vec} and neural methods~\cite{chen2020table,shraga2020web,trabelsi2020hybrid,9378239}.
It is concerned with the task of retrieving a set of tables from a table corpus and ranking them in descending order based on relevance score to a keyword query. Each table is a set of cells arranged in rows and columns like a matrix. A cell of a table could be a single word, a real number, a phrase, or even a sentence. Depending on the source of a dataset, a table has associated context fields. 
For example, the tables from Wikipedia have a caption, page title and section title, while an arbitrary Web table might have not a section title but surrounding sentences immediately before or after the table.

Researchers from the database community have been studying table search for many years~\cite{warren2006multi,xiao2009top,cafarella2008webtables,venetis2011recovering,bhagavatula2013methods,galhotra2020semantic,Zhang:2020:WTE}. 
\citet{zhang2018ad} formally re-introduced the table retrieval task and built the 1st test collection based on Wikipedia tables. 
However, there is a lack of public test collections on arbitrary Web tables for appropriate evaluation.
\citet{sun2019content} released an open domain dataset, WebQueryTable,  for table retrieval, which was collected from logs of a commercial search engine\footnote{We noticed a statistical inconsistency between released dataset and original paper.}.
\citet{shraga2020web}  constructed GNQtables dataset from a QA dataset.
Given a question-answer pair from Google Natural Questions dataset~\cite{kwiatkowski2019natural}, they extracted the table from the Wikipedia page where the answer is from and treated it as the ground truth for table retrieval.  However, this dataset is not publicly available and the details of dataset pre-processing are missing which makes it difficult to be reproduced.

\begin{figure*}[h]
\centering
\includegraphics[width=0.95\textwidth]{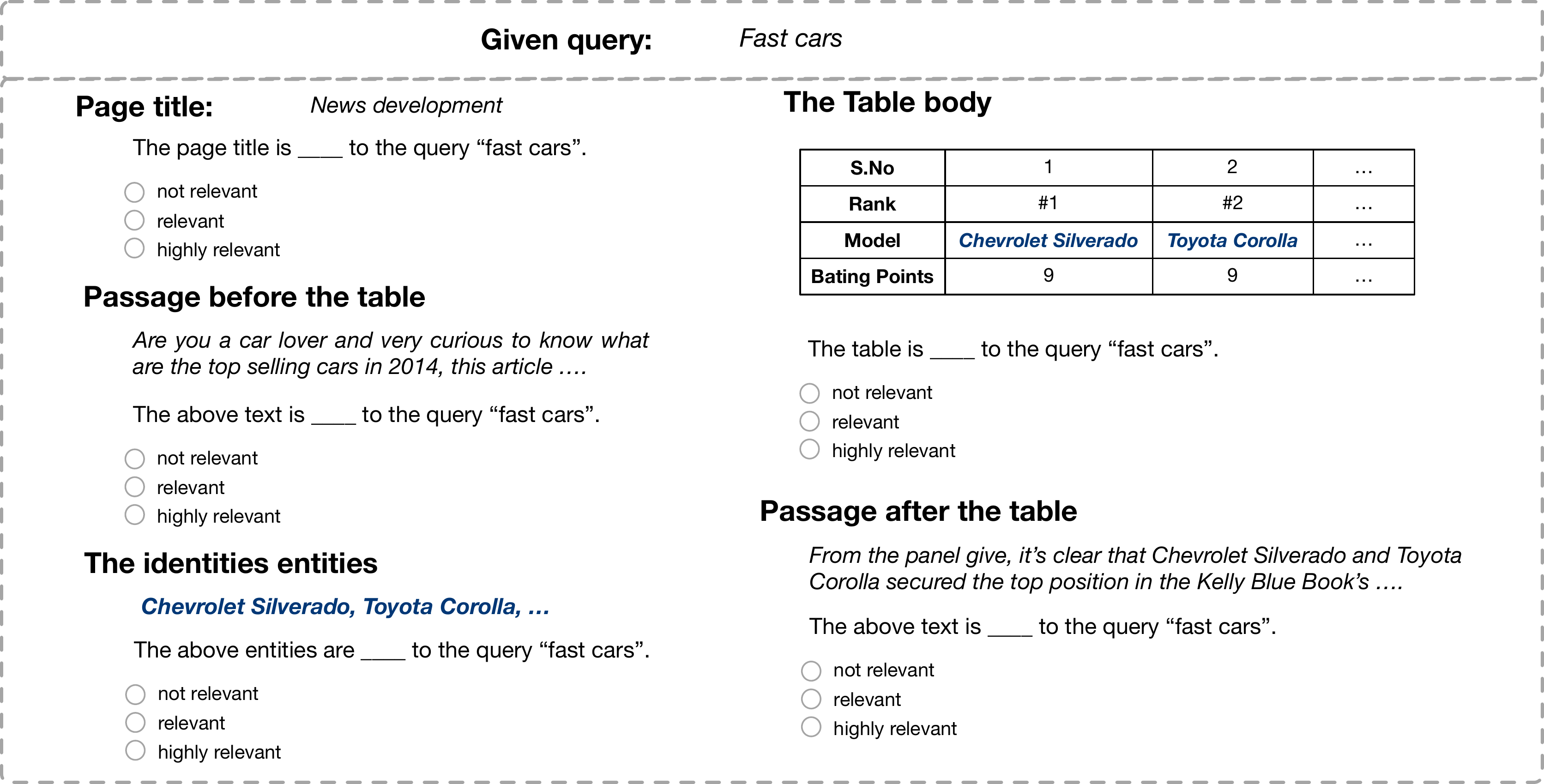}
\caption[]{Illustration of the crowdsourcing interface design.}
\label{AMT}
\end{figure*}

The main objective of this work is to create a new Web table retrieval (WTR) collection which has the following improvements compared with previous collections: 
\begin{enumerate}
    \item \textbf{Diversity.} The WikiTables collection~\cite{zhang2018ad}  and GNQtables~\cite{shraga2020web} only include tables from Wikipedia, while our collection covers broader topics (61,086 domain names).
    Our collection can include any user-generated content and less than 1\% of the tables are from Wikipedia. For the same query ``Fast cars'', the relevant tables
    in the WikiTables collection list facts about different car models, while tables in WTR can contain subjective information such as the example in Figure~\ref{AMT}\footnote{
    Wikipedia has strict content policies~(e.g., neutral point of view) and therefore is more limited to factual knowledge.}. 
    \item \textbf{Rich context.} As shown in Figure \ref{AMT}, each table in our new collection has four \textbf{context fields}: page title, text before the table, text after the table, and entities that are linked to the DBpedia~\cite{lehmann2015dbpedia}. We also keep other metadata about the source Web page. Details are in Section \ref{preprocess}.
    \item \textbf{Labels on multi-fields.} We notice that the relevance of a table and its context fields could be different. For example, the entities in Figure~\ref{AMT} are different car models and therefore can be considered as relevant to the query ``Fast cars'' while the page title ``News development'' is irrelevant. This is the first collection that has separate relevance labels for different sections of a Web page containing a table. 
    \item \textbf{Reproducibility.} We describe the details of dataset pre-processing and release the run files of baseline methods for easy comparison.
\end{enumerate}
%

The WTR collection contains not only 6,949 annotated query-table pairs but also query-context pairs for each context field which are ignored in previous test collections.
In addition, we provide details of how the corpus is pre-processed~(Section \ref{construction}) and rankings of both traditional and recently-proposed table search methods~(Section \ref{sec:exp} and \ref{result}), making a future comparison with other work easier.  
Our experimental results show that the performance of table retrieval models can be jeopardized if trained on datasets with annotation bias. With the WTR collection, we provide potential new research directions for researchers such as designing new models for table retrieval considering the relevance of context fields (Section \ref{discuss}).


\section{Constructing the Test Collection}\label{construction}

In this section, we describe the test collection, including the table corpus, queries and the process of collecting relevance assessments.

\vspace{-0.25\baselineskip}
\subsection{Query and Table Corpus}
We build the test collection with the English subset of WDC Table Corpus 2015\footnote{http://webdatacommons.org/webtables/2015/EnglishStatistics.html} which includes 50.8M relational HTML tables extracted from the July 2015 Common Crawl.
All the tables are highly relational with an indexed core column including entities or a header row describing table attributes.
\citet{zhang2020novel} further process the subset with novel entity discovery and link identified entities with their aliases to DBpedia. This results in 16.2M tables after retaining only those with at least one identified entity.
Note that the WDC Table Corpus also includes Wikipedia pages (less than 1\%) and can be considered as an extension of WikiTables collection~\cite{zhang2018ad}.
However, this new collection contains tables from 61,086 different domain names and covers a broader range of topics.

The same set of queries with WikiTables collection~ \cite{zhang2018ad} is used which includes 30 queries from~\citet{cafarella2009data} collected from Amazon's Mechanical Turk\footnote{\url{https://www.mturk.com/}} platform and 30 queries from 
\citet{venetis2011recovering} collected from query logs of searching for structured data.

\vspace{-0.25\baselineskip}
\subsection{Relevance Assessments}\label{assessment}

For our new collection, we first use a pooling method to fetch the candidate tables for all queries. Then for each query, we employ three independent annotators for relevance judgments.

\subsubsection{Pooling} \label{pool}
As a standard practice of IR test collection construction, we fetch the candidates by retrieving the top 20 tables using multiple unsupervised methods. Specifically,  we use BM25~\cite{robertson2009probabilistic} to retrieve seven indexed fields: \textit{Table}, \textit{Caption}, \textit{Page title}, \textit{Header}, \textit{TextBefore}, \textit{TextAfter} and \textit{Catchall}. The details of those fields are described in Section \ref{preprocess}.
The final assessment pool contains 6,949 query-table pairs.

\subsubsection{Collecting Relevance Judgments}
In the WikiTables collection, a table, along with other context fields such as page title and section title, is presented to an annotator, while only a single relevance label is assigned. However, a table does not always have the same relevance label as context fields. For example, the entities in Figure~\ref{AMT} representing different car models are relevant to the query ``Fast cars'' while the page title ``News development'' is irrelevant. 
Therefore, for each record (i.e., query-table pair) in our new collection, we ask the annotator to judge the relevance of each field concerning a query as shown in Figure~\ref{AMT}.  We use Amazon's Mechanical Turk to collect all the judgments based on a three-point scale:
\begin{itemize}
    \item \textbf{Irrelevant (0)}: this field
    is irrelevant to the  query (i.e., based on the context you would not expect this to
    be shown as a result from a search engine).
    \item \textbf{Relevant (1)}: this field
    provides relevant information about the query (i.e., you would expect this Web page to be included in the search results from a search engine but not among the top results).
    \item \textbf{Highly relevant (2)}: this field
    provides ideal 
    information about the query (i.e., you would expect this Web page ranked near the top of the search results).
\end{itemize}

To control the annotation quality, 135 query-table pairs for one query annotated by the authors are taken as testing questions (where at least two of the experts agreed on the relevance label). Each of the remaining query-table pairs is paired with a randomly sampled testing question, which means a single assignment consists of two query-table pairs.
Based on the testing questions, we regularly examine the following metrics: 
\begin{itemize}
    \item \textbf{Assignment accuracy}: For an assignment, if 3 out of 5 relevance judgments corresponding to gold labels of a testing question are correct, then the assignment accuracy is 60\%. 
    \item \textbf{Worker accuracy}: If a worker submits the results of 5 assignments, there are $5 \times 5=25$ labels of 5 testing questions.  If 20 out of the 25 labels are corrected, then the worker accuracy is 80\%.
    \item \textbf{Approval rate}: If a worker submits 50 assignments and 30 assignments are approved, then the approval rate of the worker is 60\%.
\end{itemize}

\begin{figure}[t]
\centering
\includegraphics[width=0.45\textwidth]{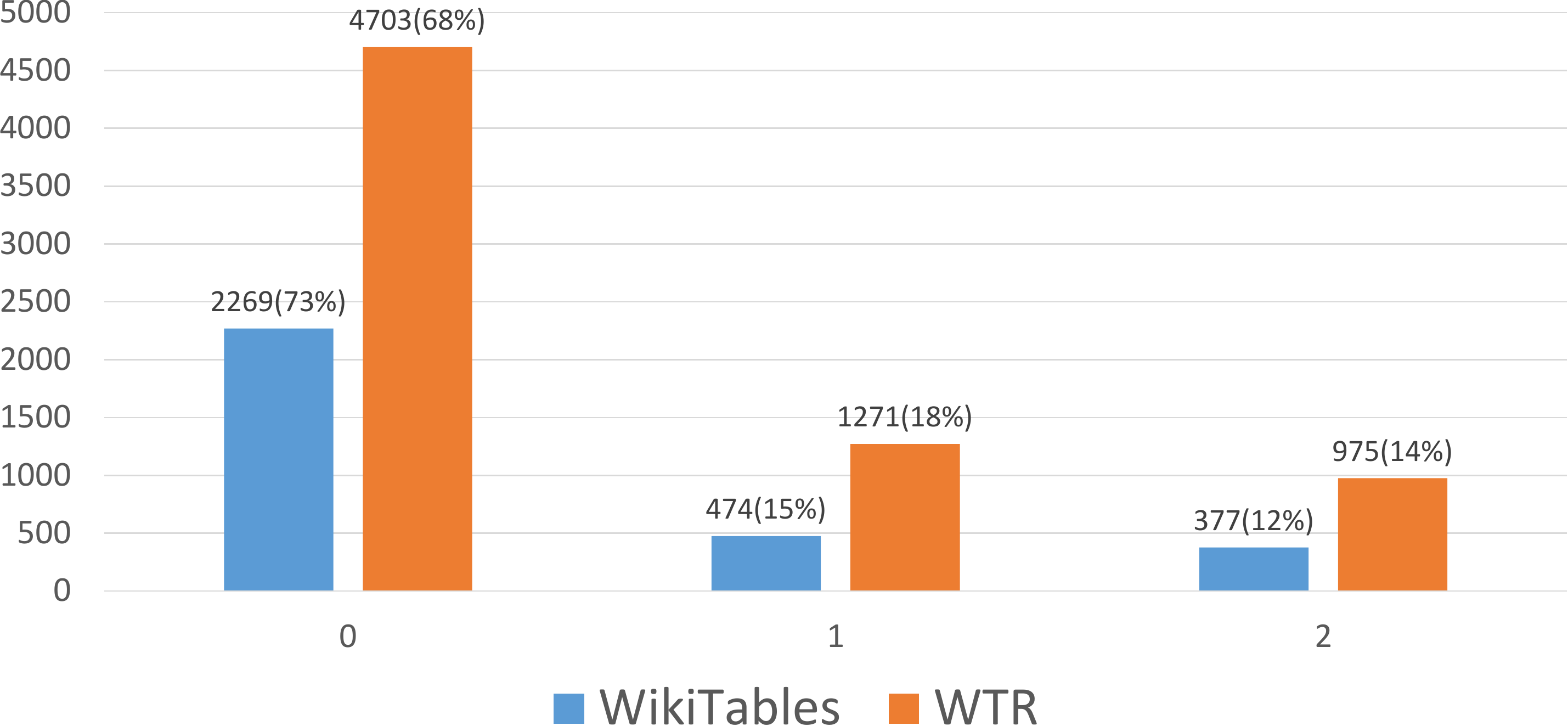}
\caption[]{Comparison of relevance labels between WikiTables and WTR collection.}
\label{label_dist}
\vspace{-0.75\baselineskip}
\end{figure}

We first approve those assignments with at least 60\% assignment accuracy or the assignments from workers whose worker accuracy is at least 80\%. Then we only allow those workers whose approval rates are at least 50\% to continue the remaining annotation process.
We collected 3 judgments for each record and paid workers 5 cents per assignment. In Table~\ref{fleiss}, we show the Fleiss' Kappa inter-annotator agreement of different sections. From the table, we can see that the annotators disagree with the judgments on entities the most and make fair agreements on other sections. 
To determine the relevance label of each section to a query, we took the majority vote among three judgments. In case of a tie, we took the average of relevance scores as the final judgment.  We compare the statistics of labels between WTR and WikiTables collection~\cite{zhang2018ad} in Figure~\ref{label_dist}. We can see that the proportion of 0,1,2 labels in the two datasets are similar but our collection is more than twice the size of the WikiTables collection. 
 
\begin{table}[b]
\vspace{-0.5\baselineskip}
\caption{The Fleiss' Kappa inter-annotator agreement of different sections.}
\label{fleiss}
\vspace{-0.5\baselineskip}
\begin{tabular}{@{}lc@{}}
\toprule
Field & Fleiss' Kappa \\ \midrule
page title & 0.28 \\
text before the table & 0.26 \\
table & 0.27 \\
entities & 0.20 \\
text after the table & 0.23 \\ \bottomrule
\end{tabular}
\end{table}

\subsubsection{Inter-field analysis}
One of our main contributions is that we provide the relevance annotations of context fields of a table. The authors of WikiTables collection~\cite{zhang2018ad} only asked workers to assign a single label for a table, but the context information (e.g., page title) was also presented to workers, which may mislead the workers.  For example, if a page title is relevant while the table is not, then a worker may still consider it as relevant. 
Besides, lack of context could lead to misunderstanding of tables. For example, the table in Figure~\ref{AMT} provides a ranked list of cars.  Without surrounding passages saying the ranking criteria are subjective ratings from customers, an annotator may think the cars in the table are ranked by speed and annotates it as relevant to the query ``fast cars''.

To study the discrepancy of relevance judgments among different fields,  we could make the assumption that the four relevance labels of context fields are also relevance judgments for the Web table.  It resembles the situation in which we ask five annotators to judge every query-table pair and we want to know the agreement between any two raters (fields). The Cohen's Kappa statistics between any two raters (fields) are summarized in Table~\ref{cross_fields}.  As we can see from the table, different context fields have different levels of agreement with \textit{Table}. Among all the context fields, \textit{TextAfter} has the most agreement with \textit{Table} while \textit{Page Title} has the least agreement with \textit{Table}. \textit{Page Title} often includes little information and even if its content is relevant the worker can hardly recognize it. However, the content in \textit{TextAfter} is more likely to be the paragraph that explains the details of the \textit{Table} which deepens the reader’s understanding. 
As shown in Figure~\ref{AMT}, \textit{Page Title} ``News Developments'' seems to be a general title of a news website and tells nothing related to the query or \textit{Table} on the same page. But the content in \textit{TextAfter} further illustrates the \textit{Table}.

\begin{table}[t]
\caption{The Cohen's kappa inter-annotator agreement between any two fields.}
\label{cross_fields}
\vspace{-0.5\baselineskip}
\resizebox{0.45\textwidth}{!}{%
\begin{tabular}{|c|c|c|c|c|c|}
\hline
           & TextBefore & Page title & Table & Entities & TextAfter \\ \hline
TextBefore & 1          & 0.44       & 0.41  & 0.34     & 0.41      \\ \hline
Page title & 0.44       & 1          & 0.39  & 0.33     & 0.39      \\ \hline
Table      & 0.41       & 0.39       & 1     & 0.4      & 0.43      \\ \hline
Entities   & 0.34       & 0.33       & 0.4   & 1        & 0.37      \\ \hline
TextAfter  & 0.41       & 0.39       & 0.43  & 0.37     & 1         \\ \hline
\end{tabular}%
}
\vspace{-0.25\baselineskip}
\end{table}

\section{Experiments}\label{sec:exp} 

In this section, we first describe the details of data pre-processing and indexing steps so that the same corpus can be reproduced. Then we present the baseline methods of table retrieval.

\subsection{Preprocessing and Indexing}\label{preprocess}

Leveraging the entity linking results of~\citet{zhang2020novel} which are formed as <table mention, entity entries> pairs\footnote{\url{https://zenodo.org/record/3627274\#.YC91NehKh1Q}}, we construct the table corpus by extracting the original tables from the English subset of WDC Table Corpus 2015\footnote{\url{http://data.dws.informatik.uni-mannheim.de/webtables/2015-07/englishCorpus/compressed/}} and mapping mentions to KB entries.
This corpus is comprised of 3M relational tables.
In addition to the original fields (cf.~Table~\ref{tbl:fields}), each table has an \emph{Entities} field, listing all the in-KB entities identified by \citep{zhang2020novel}.
We use Elasticseach\footnote{\url{https://www.elastic.co/downloads/past-releases/elasticsearch-5-3-0}} for indexing the tables, separated by the fields in Table~\ref{tbl:fields}.
\begin{table}[t]
    \centering
    \small
    \caption{Table fields used when constructing the corpus.}
    \vspace{-0.5\baselineskip}
    \begin{tabular}{l|l}
    \toprule
    \textbf{Field} & \textbf{Definition} \\
    \midrule
    \emph{Page title} &  Title of the Web page where the table lies. \\
    &It is determined from HTML tags \\
    \emph{TextBefore}& 200 words before the table \\
    \emph{Caption} & Caption of the table \\
    \emph{Table} & The extracted table from a Web page \\
    \emph{Header} &  Headers of the table if exist \\
    \emph{Entities} & Entities in the tables identified by the novel  \\
    &entity discovery process~\cite{zhang2020novel}\\
    \emph{TextAfter} & 200 words after the table \\
    \emph{Orientation} & Orientation can be either horizontal or vertical. \\
    & A horizontal table has attributes in columns while the \\
    & attributes of a vertical table are represented in rows \\ 
    \emph{URL} & The original web address of the Web page \\
    \emph{Key Column} & For a horizontal table, the key column is the one \\
    &  contains the names of the entities \\
    \emph{Catchall} & The concatenation of page title, caption, table, text\\
    & before and after the table \\ 
\bottomrule
    \end{tabular}
    \label{tbl:fields}
    \vspace{-0.5\baselineskip}
\end{table}
%
The scripts to download, preprocess and index the data are also available on our GitHub repository.\footnote{\label{repo}\url{https://github.com/Zhiyu-Chen/Web-Table-Retrieval-Benchmark}}

\vspace{-0.5\baselineskip}
\subsection{Baseline Methods} \label{method}
A line of work has developed for the ad hoc table retrieval task.
We consider the classic and some of the state-of-the-art methods as baseline approaches, which include both unsupervised and supervised methods.
\begin{itemize}
    \item \textbf{Single-field document ranking:}  Each table is represented as a single document~\citep{cafarella2008webtables}.  In our corpus, the content in the \textit{Catchall} field described in Table \ref{tbl:fields} is used as the table representation. Then classic IR methods like Language Models with Dirichlet smoothing are used for ranking.
     
    \item \textbf{Multi-field document ranking:} Each table is represented as a multi-field document~\citep{Pimplikar2012AnsweringTQ}. We use the following five fields: \textit{page title},\textit{TextBefore},\textit{Table}, \textit{TextAfter} and \textit{Header}. Unlike \citet{Pimplikar2012AnsweringTQ}  and \citet{zhang2018ad}, we do not consider \textit{Caption}. Since we find that few tables have non-empty table captions and most tables with captions are from Wikipedia.
    
    \item \textbf{LTR:} \textbf{L}earning-\textbf{T}o-\textbf{R}ank~\citep{zhang2018ad} is a feature-engineering approach leveraging table structure and lexical features (Features 1-12 in Table~\ref{tbl:features}). A random forest is used to fit the ranking features in a pointwise manner.
    
    \item  \textbf{STR:} The \textbf{S}emantic-\textbf{T}able-\textbf{R}etrieval approach~\citep{zhang2018ad} extends LTR with semantic features such as bag-of-categories, bag-of-entities, word embeddings, and graph embeddings. These embeddings are fused in different strategies~\citep{Zhang:2017:DPF} to generate ranking features (Features 13-15 in Table~\ref{tbl:features}). Like LTR, a random forest is used for pointwise regression.
    
    \item \textbf{BERT-ROW-MAX} \citet{chen2020table} propose different content selectors to select the most salient pieces of a table as BERT input. BERT-ROW-MAX, which uses BERT as the backbone and max salience selector with row items, achieves state-of-the-art performance on previous table retrieval datasets.
\end{itemize}

\textbf{M}ultimodal \textbf{T}able \textbf{R}etrieval~(MTR) ~\cite{shraga2020web} is also a recently proposed approach which learns a joint representation of the query and the different table sections. However, we find their repository\footnote{\url{https://github.com/haggair/gnqtables}} is inaccessible which leads to reproducibility difficulties. We also find their reported results of baselines (LTR, STR,etc.) on WikiTables collection are from the original runs in \cite{zhang2018ad}. It is unfair to directly compare the evaluation results since their model and baselines are not trained on the same splits used in 5-fold cross validation. Therefore, their conclusions are not convincing and we 
did not attempt to
implement their method from scratch.

\begin{table}[t]
\small
\caption{Features extracted from Web tables which are used in LTR and STR methods.}
\label{features}
\vspace{-0.5\baselineskip}
\begin{tabular}{@{}clc@{}}
\toprule
\textbf{ID} & \multicolumn{1}{c}{\textbf{Description}} & \textbf{Dim.} \\ \midrule
1 & Number of query terms & 1 \\
2 & Sum of query IDF scores (from indexed  fields except \textit{Caption}) & 6 \\
3 & The number of rows in the table & 1 \\
4 & The number of columns in the table & 1 \\
5 & The number of empty table cells & 1 \\
6 & Ratio of table size to page size & 1 \\
\begin{tabular}[c]{@{}c@{}}7\\ \  \end{tabular} & \begin{tabular}[c]{@{}l@{}}Total query term frequency \\ in the leftmost  column cells\end{tabular} & \begin{tabular}[c]{@{}c@{}}1\\ \ \end{tabular} \\
\begin{tabular}[c]{@{}c@{}}8\\ \ \end{tabular} & \begin{tabular}[c]{@{}l@{}}Total query term frequency in\\  second-to-leftmost column cells\end{tabular} & \begin{tabular}[c]{@{}c@{}}1\\ \ \end{tabular} \\
9 & Total query term frequency in the table body & 1 \\
\begin{tabular}[c]{@{}c@{}}10\\ \ \end{tabular} & \begin{tabular}[c]{@{}l@{}}Ratio of the number of query tokens found \\ in page title to total number of tokens\end{tabular} & \begin{tabular}[c]{@{}c@{}}1\\ \ \end{tabular} \\
\begin{tabular}[c]{@{}c@{}}11\\ \ \end{tabular} & \begin{tabular}[c]{@{}l@{}}Ratio of the number of query tokens found\\  in table title to total number of tokens\end{tabular} & \begin{tabular}[c]{@{}c@{}}1\\ \ \end{tabular} \\
\begin{tabular}[c]{@{}c@{}}12\\ \ \end{tabular} & \begin{tabular}[c]{@{}l@{}}Language modeling score between query and\\  multi-field document representation of the table\end{tabular} & \begin{tabular}[c]{@{}c@{}}1\\ \ \end{tabular} \\
\begin{tabular}[c]{@{}c@{}}13\\ \ \end{tabular} & \begin{tabular}[c]{@{}l@{}}Four semantic similarities between the query\\  and table represented by bag-of-word embeddings.\end{tabular} & \begin{tabular}[c]{@{}c@{}}4\\ \ \end{tabular} \\
\begin{tabular}[c]{@{}c@{}}14\\ \ \end{tabular} & \begin{tabular}[c]{@{}l@{}}Four semantic similarities between the query \\ and table represented by bag-of-entities.\end{tabular} & \begin{tabular}[c]{@{}c@{}}4\\ \ \end{tabular} \\
\begin{tabular}[c]{@{}c@{}}15\\ \ \end{tabular} & \begin{tabular}[c]{@{}l@{}}Four semantic similarities between the query\\  and table represented by bag-of-graph embeddings.\end{tabular} & \begin{tabular}[c]{@{}c@{}}4\\ \ \end{tabular} \\ \bottomrule
\end{tabular}%
\label{tbl:features}
\vspace{-0.5\baselineskip}
\end{table}
\begin{table*}[t]
\caption{The evaluation results of compared table retrieval baseline methods.}
\label{results}
\vspace{-0.5\baselineskip}
\begin{tabular}{@{}llllll@{}}
\toprule
\textbf{Model} & \textbf{MAP} & \textbf{P@5} & \textbf{P@10} & \textbf{NDCG@5} & \textbf{NDCG@10} \\ \midrule
Single-field document ranking & 0.5071 & 0.4380 & 0.3966 & 0.4124 & 0.4851 \\
Multi-field document ranking & 0.5104 & 0.4407 & 0.3943 & 0.4201 & 0.4916 \\
LTR & 0.5878 & 0.5300 & 0.4433 & 0.5313 & 0.5870 \\
STR & 0.6210 & 0.5493 & 0.4727 & 0.5585 & 0.6258 \\
BERT-ROW-MAX(base) & \textbf{0.6346} & \textbf{0.5713} & \textbf{0.4800} & \textbf{0.5737} & \textbf{0.6327} \\ \bottomrule
\end{tabular}
\end{table*}
\begin{table*}
\caption{Table retrieval evaluation results with different annotation ``bias''. We highlight the cases where the results are improved over the original method.}
\label{adv}
\vspace{-0.5\baselineskip}
\begin{tabular}{@{}llllll@{}}
\toprule
\multicolumn{1}{l}{\textbf{Model}} & \multicolumn{1}{l}{\textbf{MAP}}  & \multicolumn{1}{l}{\textbf{P@5}}  & \multicolumn{1}{l}{\textbf{P@10}} & \multicolumn{1}{l}{\textbf{NDCG@5}} & \multicolumn{1}{l}{\textbf{NDCG@10}} \\ \midrule
STR                       & 0.6210                   & 0.5493                   & 0.4727                   & 0.5585                     & 0.6258                      \\
STR (Max-Entities)         & \textbf{0.6261 (+0.82\%)} & \textbf{0.5573 (+1.46\%)} & \textbf{0.4763 (+0.77\%)} & \textbf{0.5634 (+0.88\%)}   & \textbf{0.6281 (+0.36\%)}    \\
STR (Min-Entities)         & 0.5893 (-5.11\%)          & 0.5253 (-4.37\%)          & 0.4533 (-4.10\%)          & 0.5251 (-5.99\%)            & 0.5872 (-6.18\%)             \\
STR (Max-PageTitle)        & 0.6137 (-1.18\%)          & 0.5487 (-0.12\%)          & 0.4677 (-1.06\%)          & 0.5517 (-1.22\%)            & 0.6154 (-1.67\%)             \\
STR (Min-PageTitle)        & 0.6060 (-2.42\%)          & 0.5473 (-0.36\%)          & 0.4613 (-2.40\%)          & 0.5419 (-2.98\%)            & 0.6041 (-3.47\%)             \\
STR (Max-TextBefore)       & 0.6127 (-1.33\%)          & 0.5380 (-2.06\%)          & 0.4703 (-0.50\%)          & 0.5472 (-2.03\%)            & 0.6171 (-1.39\%)             \\
STR (Min-TextBefore)       & 0.6021 (-3.04\%)          & 0.5480 (-0.24\%)          & 0.4547 (-3.81\%)          & 0.5471 (-2.04\%)            & 0.6003 (-4.08\%)             \\
STR (Max-TextAfter)        & 0.6203 (-0.12\%)          & \textbf{0.5533 (+0.73\%)} & \textbf{0.4770 (+0.91\%)} & 0.5545 (-0.72\%)            & \textbf{0.6265 (+0.11\%)}    \\
STR (Min-TextAfter)        & 0.6007 (-3.26\%)          & 0.5327 (-3.03\%)          & 0.4577 (-3.17\%)          & 0.5294 (-5.21\%)            & 0.5970 (-4.61\%)             \\ \bottomrule
\end{tabular}
\end{table*}
\subsection{Implementation Details}
For single-field and multi-field document ranking, we use the implementations from Nordlys~\cite{hasibi2017nordlys} which has interfaces to Elasticsearch. 
Since the features used by LTR rely on the source of tables and some features extracted from Wikipedia tables are not available for Web tables (e.g., number of page views), we generate the features which apply to the new test collection.  The original STR calculates four semantic representations for queries and tables. Since there are no Wikipedia categories for all Web tables, we use three semantic representations for queries/tables: bag-of-entities, word embeddings, and graph embeddings.  For each type of semantic representation, we use early fusion, late-max, late-sum, and late-max, as described in \citet{Zhang:2017:DPF} to obtain four semantic matching scores, which results in 12 semantic matching features in total for each query-table pair.
We summarize the features in Table~\ref{features}, where features 1-12 are used in LTR and features 1-15 are used in STR. The scikit-learn\footnote{https://scikit-learn.org/} implementation of random forest is used for both LTR and STR. We set the number of trees to 1000 and a maximum number of features in each tree to 3. For BERT-ROW-MAX, we use the pre-trained BERT-base-cased model\footnote{\url{https://huggingface.co/transformers/pretrained_models}.html} which consists of 12 layers of Transformer blocks. As in \citet{chen2020table}, we train the model with 5 epochs, and batch size is set to 16.  The Adam optimizer~\cite{Kingma2015AdamAM} with a learning rate of 1e-5 is used to optimize BERT-ROW-MAX. A linear learning rate decay schedule with a warm-up ratio of 0.1 is also used.
We train all the models using 5-fold cross-validation (w.r.t. NDCG@5). To facilitate fair comparison in future work, we prepared the five data splits used for cross-validation in our WTR repository\textsuperscript{\ref{repo}}, which is not provided by previous work~\cite{zhang2018ad}.

\section{Results and Analysis}\label{result}

\subsection{Overall Performance}
In Table~\ref{results} we report on the performance of different table retrieval methods. The conclusion is consistent with previous work~\cite{zhang2018ad,chen2020table} that BERT-ROW-MAX achieves the best overall evaluation \mbox{metrics}. STR outperforms LTR and other unsupervised methods significantly. However, different from the results on WikiTables collection, BERT-ROW-MAX does not outperform STR sufficiently. We speculate that the new corpus may include more noise which makes it more difficult for neural models to learn features. In contrast, LTR and STR are trained on curated features which are more robust to noise in raw text.

\vspace{-0.5\baselineskip}
\subsection{Utilizing Labels of Context Fields}
\label{sec:results:uml}

Recall that we have explicitly collected the relevance judgments of different fields in Section~\ref{assessment}, we observe that a context field does not always have the same relevance label as the table. 
This discrepancy could result in bias in the annotation process of WikiTables collection. For example, a strict annotator may think a table should be annotated as relevant when all the context fields are also relevant. While a lax annotator may consider a table as relevant when any of the context fields provides related information even if the table itself is not relevant.  
To investigate how the labels of context fields can help table retrieval, e.g., how the annotation bias could potentially affect the models, we employ two strategies to resemble a strict annotator and a lax annotator: given the original label $l_t$ of a table and the label $l_c$ of one context field, we re-annotate $l_t$ as either $max(l_t,l_c)$ or $min(l_t,l_c)$. Then we can use the ``new'' labels to train the models but evaluate on original labels.

We take the STR approach as the example and in total form 8 variations for STR which are trained with ``biased'' labels by combining the min/max strategy with four context fields. 
We summarize the results of STR trained on new labels in Table~\ref{adv}.  For example, STR (max-entity) means STR is trained with the new labels where the maximum relevance among a table and its \textit{Entities} field is used as the training label for each query-table pair. Note that we only change the labels in the training set and keep the original testing set since the task is still table retrieval rather than context field retrieval.
We can observe that the performance of majority models decreases compared with the model trained on the original labels, which demonstrates that the biased annotation can harm the model training. Among those cases, STR (Min-Entities) has the most performance drop.
Nevertheless, there are a few cases where the performance slightly increased. It is worth noting that STR (max-entity) performs better than the model trained on the original labels on all the evaluation metrics. We observe the same results on other supervised methods. 
Table retrieval and entity retrieval may be highly synergistic tasks. Taking good advantage of labels from \textit{Entities} can help table retrieval while misusing them can lead to a large performance drop.

\section{Discussion}\label{discuss}

\subsection{Availability}

We publish the following resources in our WTR repository\textsuperscript{\ref{repo}}:
\begin{itemize}
    \item \textbf{Scripts}: for reproducibility, we release the scripts to download and preprocess the WDC Table corpus, and match the entity linking results in \cite{zhang2020novel}.
    \item \textbf{WTR table dump}: for ease of use, we release the pre-processed table corpus used for indexing and the pooling results used for relevance judgments.
    \item \textbf{Annotations}: we provide separate files containing the relevance judgments for query-table and query-context pairs. We also provide the five splits used in this paper for cross-validation. 
    \item \textbf{Baselines}: we release the run files of baselines presented in this paper. For STR and LTR, we also release the reproduced features.
\end{itemize}

\subsection{Future Directions}

We propose the following research questions to indicate future research directions on utilizing this new test collection.

\begin{itemize}
\setlength{\itemindent}{-1.8em}
    \item \textbf{RQ1}: \emph{Do current table retrieval methods have good transferability?}
\end{itemize}

The WTR collection is complementary to the previous WikiTables collection~\cite{zhang2018ad} and covers a broader range of topics. 
Wikipedia pages usually have a similar structure and the tables are well-formatted. However, the structure of different websites can be very different. For example, the page title of Wikipedia usually represents an entity or a very informative event.  While for some websites, the page title can be very general and does not describe anything specific about the Web page (e.g., example in Figure~\ref{AMT}).
Therefore, it is possible that some features extracted from one domain do not generalize well to another domain.  For example, when we reproduce the features of LTR and STR, we disregard some features proposed for Wikipedia specifically. It is an interesting question to study the transferability of different models and what features extracted from one dataset (training set) also generalize to another dataset (testing set). 

\begin{itemize}
\setlength{\itemindent}{-1.8em}
\item \textbf{RQ2}: \emph{How can the relevance signals from context fields further improve the table retrieval task?}
\end{itemize}

 In our experiments, we find that with a simple strategy to combine the relevance labels of the table and its \textit{Entities} field, the model performance can increase (cf.~Sect.~\ref{sec:results:uml}). A natural follow-up question is: how can we design a model that automatically utilizes the labels of context fields to help table retrieval? 
This task is similar to multi-field document retrieval~\cite{zamani2018neural,robertson2004simple}  where the models take the advantage of document structure and handle multiple document fields.  In the setting of Web table retrieval, multiple context fields can be created even when they are not tagged in the HTML source code. \citet{zamani2018neural} treat the body of a page as a single field, while for table search the body text can be further split into the passages before and after the table. Though previous works in multi-field document retrieval utilize all fields, the field-level relevance is not considered due to the lack of such annotations. Therefore, the WTR dataset can also be used as a collection for multi-field document retrieval. 

\begin{itemize}
\setlength{\itemindent}{-1.8em}
\item \textbf{RQ3}: \emph{Could other tasks benefit table retrieval or vice versa?}
\end{itemize}

Though the objective of table retrieval is to return a list of relevant tables given a user query, a table alone may not satisfy a user's information need. 
For example, the table in Figure \ref{AMT} provides a ranked list of cars but it does not tell how the scores are rated. Given the query ``fast cars'', a user may think the cars in the table are ranked by speed.  However, the surrounding passages say the ranking criteria are subjective ratings from customers. Without any explanation about the table, a user could misunderstand the semantic meanings in the table. Therefore, context information is very important for a user to understand the table accurately. If we also consider passage retrieval as an auxiliary task for table retrieval, the search results presented to users can be more understandable. 

In addition, Web table retrieval is also relevant to the task of entity retrieval~\cite{chen2016empirical,balog2013test,hasibi2017dbpedia}. Our test collection includes a large number of entities extracted from tables and linked to DBpedia. If a table is relevant, then it is very likely that the entities in the table are also relevant. Therefore, a model trained for table retrieval may also help entity retrieval.  Our experimental results in Section~\ref{result} also indicate that table retrieval and entity retrieval are synergistic.
The recent signs of progress in multi-task learning~\cite{Liu2019MultiTaskDN,standley2020tasks} and pretraining techniques ~\cite{Devlin2019BERTPO,Peters2019KnowledgeEC} have shown that a model pre-trained or jointly trained on related tasks can help learn more generalized features and improve the model performance on target tasks. It will be interesting to study how we can utilize the datasets for entity retrieval~\cite{balog2013test,hasibi2017dbpedia} to train models for table retrieval, or use the WTR collection to help train models for entity search.

\section{Conclusion}

In this work, we describe a new test collection WTR for the task of table retrieval. Compared with previous datasets, WTR covers a broader range of topics and includes tables from over 61,000 different domain names. We present a detailed walk-through of how the test collection was created. In addition to the WTR collection, we offer the runs of baseline methods and feature sets used in previous work.
We also discussed a number of interesting opportunities for researchers to use WTR in their future work.
\subsection*{Acknowledgments}
This material is based upon work supported by the National
Science Foundation under Grant No.\ IIS-1816325. We thank the anonymous reviewers for their insightful comments.

\bibliographystyle{ACM-Reference-Format}
\bibliography{acmart.bib}
\end{document}